%% file: main.tex
\documentclass[aps,prl,noeprint,twocolumn,showpacs,byrevtex,amsmath,amssymb]{revtex4-2}
\usepackage{amsmath}
\usepackage{graphicx}
\usepackage{subfigure}
\usepackage{epstopdf}
\usepackage{epsfig}
\usepackage{color}
\usepackage{multirow}
\usepackage{setspace}
\usepackage{overpic}
\usepackage{amssymb}
\usepackage{units}
\usepackage{lineno}
\usepackage{bm}
\usepackage{rotating}
\usepackage[utf8]{inputenc}
\usepackage[bookmarksnumbered,pdfstartview=FitH,colorlinks,urlcolor=blue,citecolor=blue,linkcolor=blue]{hyperref}
\let\oldequation\equation
\let\oldendequation\endequation

\renewenvironment{equation}
{\linenomathNonumbers\oldequation}
{\oldendequation\endlinenomath}

\newcommand{\ifb}{{\rm{fb}}^{-1}}
\newcommand{\gev}{\,\unit{GeV}}

\newcommand{\mbc}{M_{\rm{BC}}}
\newcommand{\Lc}{\Lambda_c^{+}}
\newcommand{\ALc}{\bar{\Lambda}_c^{-}}
\newcommand{\Lcppi}{\Lambda_c^{+}\to p\pi^0}
\newcommand{\Lcpeta}{\Lambda_c^{+}\to p\eta}

\begin{document}

\oddsidemargin  -0.2cm
\evensidemargin -0.2cm

\title{\bf \boldmath
    Observation of the Singly Cabibbo-Suppressed Decay \texorpdfstring{$\Lcppi$}{Lambdac->p pi0}}

\input{authorlist_2024-01-27}

\begin{abstract}
    Utilizing 4.5~$\ifb$ of $e^+e^-$ annihilation data collected with the BESIII detector at the BEPCII collider at center-of-mass energies between 4.600 and 4.699 GeV, the first observation of the singly Cabibbo-suppressed decay $\Lcppi$ is presented, with a statistical significance of $5.4\sigma$. The ratio of the branching fractions of $\Lcppi$ and $\Lcpeta$ is measured as $\mathcal{B}(\Lcppi)/\mathcal{B}(\Lcpeta)=(0.120\pm0.026_{\rm stat.}\pm0.007_{\rm syst.})$. This result resolves the longstanding discrepancy between earlier experimental searches, providing both a decisive conclusion and valuable input for QCD-inspired theoretical models. A sophisticated deep learning approach using a Transformer-based architecture is employed to distinguish the signal from the prevalent hadronic backgrounds, complemented by thorough validation and systematic uncertainty quantification.
\end{abstract}

\maketitle

Weak decays of charmed hadrons provide profound insights into the intricate interplay between the strong and weak interactions~\cite{Cheng:2021qpd}. The decay amplitudes typically consist of factorizable and non-factorizable components~\cite{Chau:1982da}. In the case of charmed mesons, non-factorizable contributions are often considered negligible~\cite{Chau:1986jb}. However, this approximation does not hold for charm baryons~\cite{Chau:1995gk}. For instance, there is a significant impact from non-factorizable $W$-exchange diagrams, which evade both helicity and color suppression in the charmed baryon sector~\cite{Cheng:1993gf}.
In the two-body singly-Cabibbo-suppressed (SCS) decay $\Lcppi$, a mixture of factorizable $W$-emission process with non-factorizable $W$-emission and $W$-exchange processes is encountered. This complex scenario challenges our comprehension of charm decay dynamics~\cite{Kohara:1998jm}.

The last decade witnessed rapid progresses in studies of charm baryon weak decays, both theoretically and experimentally~\cite{Ke:2023qzc,Li:2021iwf}. Various phenomenological models and approaches have been explored for $\Lcppi$, including constituent quark model~\cite{Uppal:1994pt}, heavy quark effective theory~\cite{Chen:2002jr}, dynamical model calculations~\cite{Cheng:2018hwl,Zou:2019kzq}, topological diagrams~\cite{Zhao:2018mov,Hsiao:2021nsc}, and SU(3) flavor symmetry~\cite{Sharma:1996sc,Lu:2016ogy,Geng:2018plk,Geng:2018rse,Geng:2019xbo,Xing:2023dni}. Nevertheless, the precision of theoretical predictions is overall limited owing to the above complexities and the scarcity of experimental input. Recently, the BESIII collaboration reported the first evidence for the decay $\Lcppi$ with a statistical significance of $3.7\sigma$, and measured its branching fraction (BF) to be $(1.56^{+0.72}_{-0.58}\pm0.20)\times10^{-4}$~\cite{BESIII:2023uvs}. Notably, this result is inconsistent with the previously established upper limit of $0.8\times10^{-4}$ by the Belle experiment~\cite{Belle:2021mvw}. A definitive observation is crucial to resolve this discrepancy and calibrate the theoretical predictions.

The primary challenge in measuring $\Lcppi$ lies in distinguishing the signal from the substantial hadronic background produced in $e^+e^-$ annihilation. Utilizing the unique near-threshold production of $\Lc\ALc$ pairs at BESIII, Ref.~\cite{BESIII:2023uvs} employed a ``double-tag'' strategy~\cite{MARK-III:1985hbd,Li:2021iwf} that searched for the signal mode in $\Lc$ decays while simultaneously reconstructing the $\ALc$ in other selected decay modes. Although this strategy effectively reduces the hadronic background, it suffers from considerable efficiency loss due to the required exclusive reconstruction of the selected subset of $\ALc$ decays. In contrast, Ref.~\cite{Belle:2021mvw} opted to directly reconstruct $\Lcppi$ without imposing constraints on the rest of the $e^+e^-$ annihilation event. As a result, potential signals become obscured by a high background level.

To address the trade-off between signal efficiency and background level, we resort to a deep neural network (DNN) which has exhibited remarkable capabilities for uncovering new relations and hidden patterns, showing promise in many data processing fields~\cite{deeplearning}, including particle physics~\cite{Schwartz2021Modern}. Compared to selection-based methods, the topological characteristic of $e^+e^-$ annihilation events can be efficiently recognized and interpreted by a DNN. By training a DNN with all possible $\ALc$ decay modes, we can differentiate between events containing a $\ALc$ and those without, achieving the same goal as the double-tag technique but with higher efficiency. This approach parallels recent advancements of jet tagging in LHC experiments~\cite{Larkoski:2017jix,Kogler:2018hem}, but at a lower energy scale. To our knowledge, this is one of the first applications of such a technique in electron-positron collider experiments~\cite{hepmllivingreview}.

In this Letter, we present the first observation of the SCS decay $\Lcppi$ using 4.5~$\ifb$ of $e^+e^-$ collision data collected with the BESIII detector at seven center-of-mass (c.m.) energies between 4.600 and $4.699\gev$~\cite{BESIII:2022dxl,BESIII:2022ulv}. We employ a single-tag strategy by reconstructing one $\Lc$ baryon in its decay to $\Lcppi$. A DNN is then invoked for binary classification of signal and background in the rest of the event. Apart from $\Lcppi$, the decay of $\Lcpeta$ is also studied as a reference channel. The ratio of their BFs is reported in order to offset some systematic uncertainties, especially those related to the DNN model. Throughout this Letter, charge conjugate modes are implied.


Details about the BESIII detector design and performance are provided in Ref.~\cite{BESIII:2009fln}. An inclusive Monte Carlo (MC) simulated data sample, as described in Ref.~\cite{BESIII:2022bkj}, is employed to estimate backgrounds. A signal MC sample of $e^+e^-\to\Lc\ALc$ is generated to determine detection efficiencies, where one $\Lambda_{c}$ decays to the signal (reference) state and the $\pi^0$ ($\eta$) decays subsequently into two photons, while the other $\Lambda_{c}$ in the event decays to all possible final states. Both the inclusive and signal MC samples are used in training the DNN.

The selections of $\Lcppi$ and $\Lcpeta$ decays follow the previous BESIII studies~\cite{BESIII:2017fim,BESIII:2023ooh}, with two modifications in the selection of $\pi^0$($\eta$). Photon candidates are required to have an opening angle of larger than 10 degrees, measured at the interaction point, with respect to any charged tracks, except for those identified as anti-protons where the angle must be larger than 20 degrees. The veto on the decay angle of one photon in the $\pi^0$($\eta$) rest frame is not imposed, as the DNN is expected to improve the signal-to-background ratio. A kinematic observable, the beam-constrained mass $\mbc$, is utilized for signal extraction. Here, $\mbc=\sqrt{E^2_{\rm beam}/c^4-|\vec{p}_{\Lc}|^2/c^2}$, where $E_{\rm beam}$ is the beam energy and $\vec{p}_{\Lc}$ is the momentum of the $\Lc$ candidate in the $e^+e^-$ rest frame. 
As illustrated in Fig.~\ref{fig:comp}, the signal of $\Lcpeta$ is discernible against the backgrounds, while the high background level impedes the observation of the signals of $\Lcppi$. These backgrounds primarily comprise hadronic final states produced from $e^+e^-\to q\bar{q}\,(q=u,\,d,\,s)$, while the contributions from other non-signal $e^+e^-\to\Lc\ALc$ events are almost negligible.

\begin{figure}[htbp]
    \centering
    \includegraphics[width=0.4\textwidth]{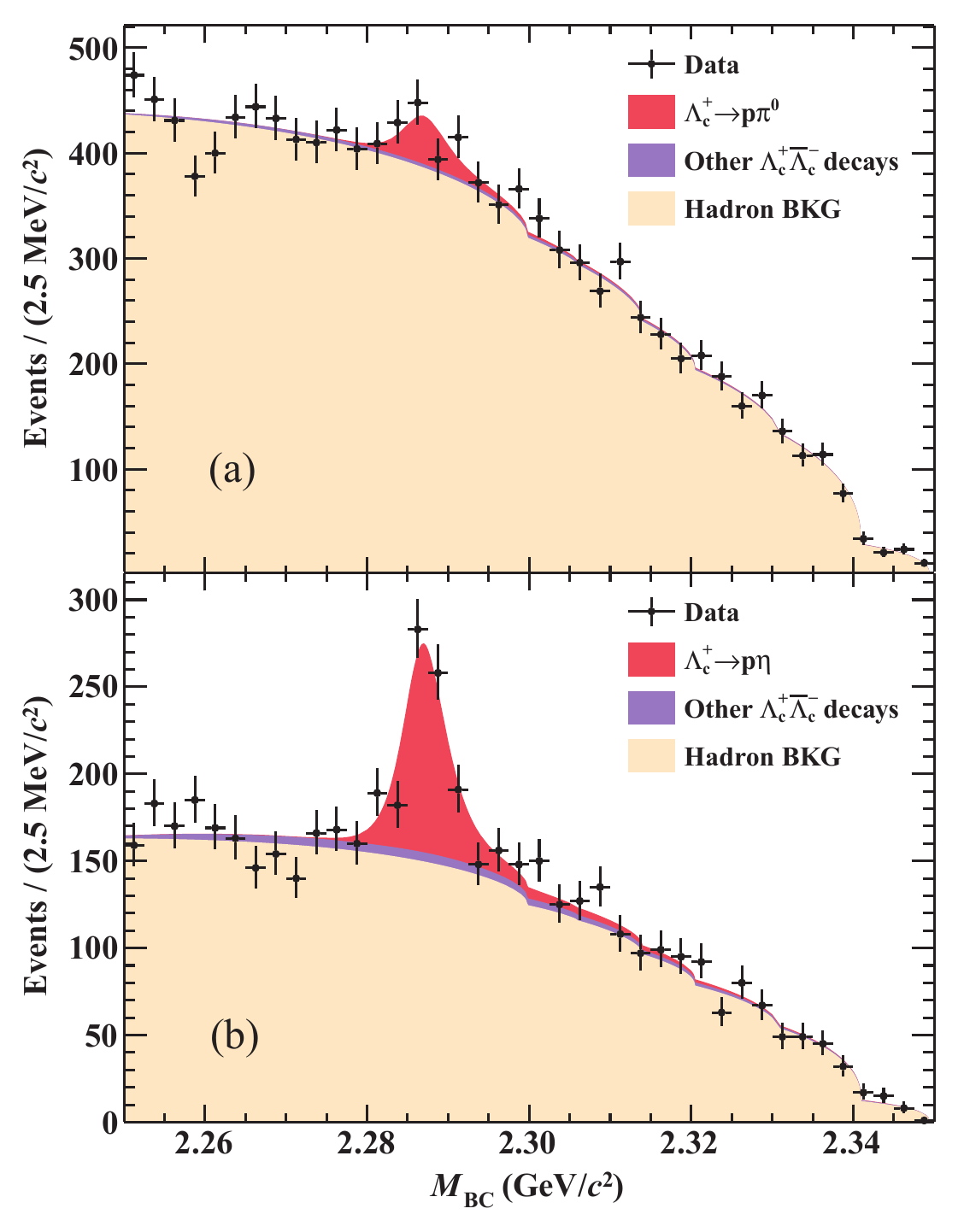}
    \caption{Distributions of $\mbc$ for (a) $\Lcppi$ and (b) $\Lcpeta$ before implementing deep learning. The black dots with error bars represent data, and the colored areas indicate the MC-simulated contributions of the signal and two background components.}
    \label{fig:comp}
\end{figure}


In particle physics experiments, deep learning approaches aim to process large sets of minimally refined data at the level of fundamental detector responses. It has proven to be more effective than traditional multivariate techniques~\cite{Voss:2009rK} like boosted decision trees, with jet tagging being a leading example~\cite{Larkoski:2017jix,Kogler:2018hem}. Among the evolving deep learning algorithms~\cite{deOliveira:2015xxd,Guest:2016iqz,Louppe:2017ipp,henrion2017neural,Komiske:2018cqr} for jet tagging, the {\sc Particle Transformer} (ParT)~\cite{qu2022particle} stands out for its state-of-the-art performance. The jet components are conceptualized as a {\it particle cloud}~\cite{Qu:2019gqs}, i.e., an unordered and variable-sized set of the outgoing particles, where each particle is characterized by its spatial coordinates and additional features such as type, momentum, etc. A model architecture inherited from the Transformer~\cite{vaswani2017attention} is leveraged to process this data representation. For this study, ParT is chosen as the basis of our DNN considering its impressive capabilities, and several adaptations are made to tailor it for the BESIII experiment, as elaborated below.

The dataset for training DNN is prepared as a random shuffle of signal and background events, with equal statistics after signal selection. The signal events are from the signal MC samples, while the background events are drawn from the inclusive MC sample excluding the signal process. Events from different c.m.~energies are included in proportion to their selected yields in the real data. The dataset comprises approximately 2.3 million events for $\Lcppi$ and 1.1 million for $\Lcpeta$. From these, 80\% of the events are allocated for actual training, and the remaining 20\% are reserved for independent validation. It is noted that a combined dataset of $\Lcppi$ and $\Lcpeta$ is fed into one unified DNN, aiming to generalize the performance across both decay modes as a prerequisite for using $\Lcpeta$ for reference. An additional treatment is employed on the dataset to address potential correlations between the DNN output and the mass spectra. This correlation could inadvertently create signal-like structures in the mass spectrum of remaining background events not rejected by the DNN output, an effect known as {\it mass sculpting}~\cite{Dolen:2016kst,Englert:2018cfo,Kasieczka:2020yyl,Kitouni:2020xgb}. Inspired by the iterative weighting method~\cite{Sun:2020ehv} in the implementation of initial-state radiation corrections in cross-section measurements, we establish a similar approach to decouple the DNN output from the $\mbc$ observable. A weight, $\omega(\mbc)$, acting on the loss function of the DNN is assigned to each background event. Events with higher weights contribute more to the loss function, thus being more effectively classified during training. The weight $\omega_i^j(\mbc^j)$ for event $j$ in the $i^{th}$ iteration is calculated as
\begin{equation}
    \omega_i^j(\mbc^j)=\omega_{i-1}^j(\mbc^j) \, \left[ \left.\frac{p^{\rm BKG}_{i-1}(\mbc)}{p^{\rm BKG}_{\rm orig}(\mbc)}\right|_{\mbc=\mbc^j} \right],
    \label{eq:planing}
\end{equation}
where $\omega_0^j(\mbc^j)=1$, $p^{\rm BKG}_{i-1}(\mbc)$ represents the normalized probability density function for the remaining background shape in the $(i-1)^{th}$ iteration, and $p^{\rm BKG}_{\rm orig}(\mbc)$ denotes that for the original background shape. With each iteration, the DNN is retrained, and both $p^{\rm BKG}_i(\mbc)$ and $\omega_i^j(\mbc^j)$ are updated. Our findings indicate that convergence is achieved when $p^{\rm BKG}_i(\mbc)$ approximates $p^{\rm BKG}_{\rm orig}(\mbc)$, as shown in Fig.~\ref{fig:iter}, resulting in a trivial background shape that facilitates determination of the signal yield.

\begin{figure}[htbp]
    \centering
    \setlength{\abovecaptionskip}{-0.2ex}
    \includegraphics[width=0.4\textwidth]{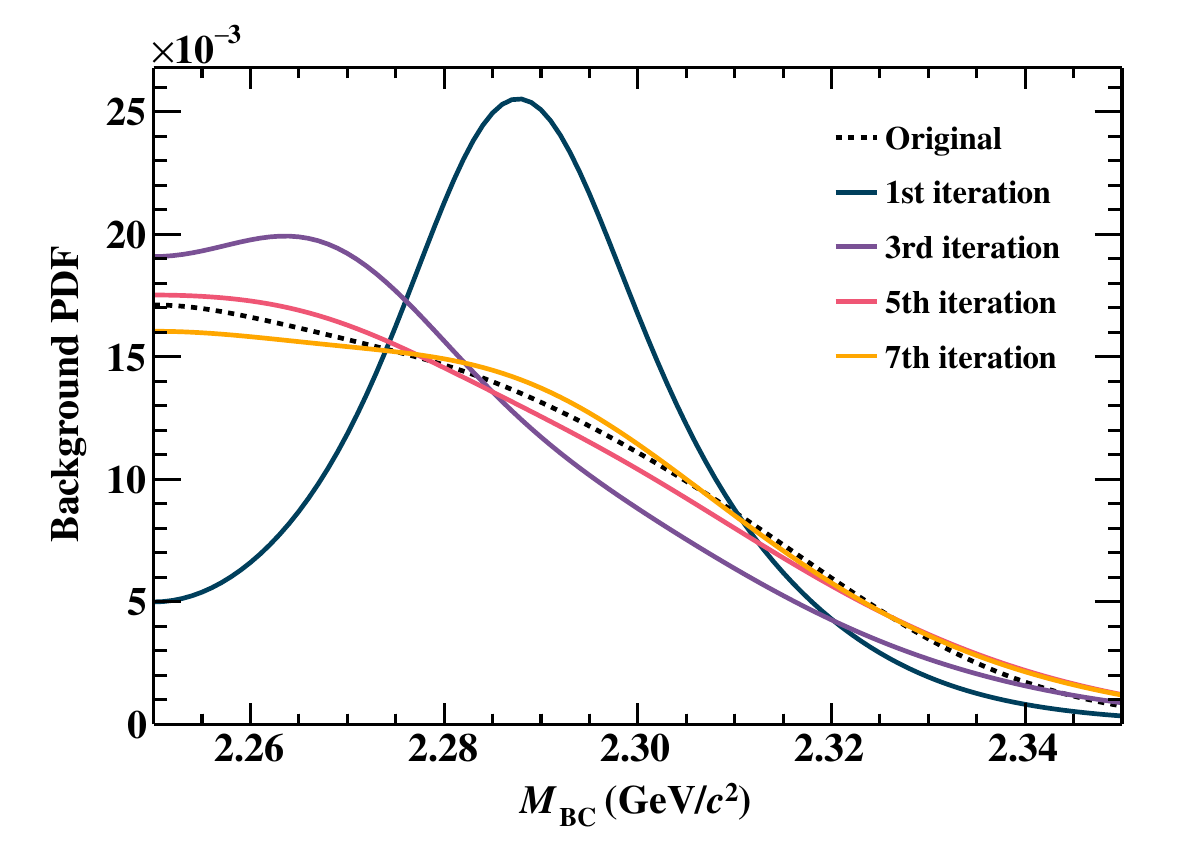}
    \caption{Normalized probability density function (PDF) in $\mbc$ for $\Lcppi$ background shapes during the iteration process.}
    \label{fig:iter}
\end{figure}

The input information from an event to the DNN includes all charged tracks reconstructed in the main drift chamber (MDC) and isolated showers clustered in the electromagnetic calorimeter (EMC), which creates two sets of {\it particle clouds}. For every charged track, the data used includes the azimuthal and polar angles in the laboratory frame, charge, the magnitude of momentum, and parameters characterizing the helical trajectory in the MDC. Additionally, normalized differences between the measured and predicted specific energy loss ${\rm d}E/{\rm d}x$ for the electron, proton, pion and kaon hypotheses, along with similar differences for the time-of-flight (TOF) are used to summarize particle identification information. For each shower, the properties include the azimuthal and polar angles, the energy deposition, the number of crystals above the threshold, the shower time, the energy ratio between $3\times3$ and $5\times5$ crystal groupings surrounding the shower center, the lateral and secondary moments, and the A20 and A42 Zernike moments~\cite{Sinkus:1996ch}. Data and MC simulations are overall consistent for these adopted features.

The model architecture of the DNN is a streamlined version of ParT relative to its original form~\cite{qu2022particle}, given the simpler topology of $e^+e^-$ annihilation events at BESIII compared to the jet substructures at the LHC. This adaptation includes three particle attention blocks and one class attention block. It uses a particle embedding encoded from the input particle features using a 3-layer MLP with $(128,\ 256,\ 128)$ nodes in each layer, and an interaction embedding encoded from the input pairwise features using a 4-layer point-wise 1D convolution with $(64,\ 64,\ 64,\ 8)$ channels. The training is performed on the aforementioned dataset for 75 epochs with a batch size of 512 and an initial learning rate of 0.001. These hyper-parameters have been optimized to maximize the performance of DNN while preventing overfitting. All other settings are consistent with those detailed in the original ParT publication. In addition, a machine-learning technique called {\it model ensemble} is employed to act on the model. With some randomness factors incorporated in the training such as network initialization, batch processing sequence and dropout~\cite{Srivastava:2014kpo} mechanism, a total of 20 DNNs are trained in parallel. By averaging their outputs for each event at inference, we create an ensemble DNN that exhibits greater robustness and generalization ability than the individual trained DNNs.

The output of the DNN is a score between $[0,\,1]$ assigned to each event, reflecting its probability of belonging to the signal category. We therefore require the score to exceed 0.95 as the final step of event selection. This threshold has been optimized by maximizing the figure-of-merit $\frac{S}{\sqrt{S+B}} \times \frac{S}{S+B}$ for $\Lcpeta$, where $S$ ($B$) denotes the expected signal (background) yield from a MC study with data-matched luminosities. Figure~\ref{fig:fit} illustrates the $\mbc$ spectra for $\Lcppi$ and $\Lcpeta$ after the deep learning implementation. The backgrounds are significantly reduced in both channels, to approximately down to 1/20 of their original levels, with a relative signal loss of about 40\%; this enables the observation of the signal for $\Lcppi$.

The relative BF ratio $\mathcal{B}(\Lcppi)/\mathcal{B}(\Lcpeta)$ is derived from unbinned maximum likelihood fits to their $\mbc$ spectra. These fits are simultaneously conducted on both decay channels and across various c.m.~energies. In each fit, the signal shape is modeled with an MC-simulated profile convolved with a Gaussian function to account for resolution discrepancies. The parameters of the Gaussian function are shared between the $\Lcppi$ and $\Lcpeta$ fits due to the limited statistics of the former mode. The background is described by an ARGUS function~\cite{ARGUS:1990hfq}, given that no peaking structure is indicated by the MC study. A small contribution from non-signal $e^+e^-\to\Lambda_c^+\bar{\Lambda}_c^-$ decays is included using a fixed yield and the MC-simulated shape. The ratio of BFs is shared among each fit with the relation
\begin{equation}
    \frac{\mathcal{B}(\Lcppi)}{\mathcal{B}(\Lcpeta)}=\frac{\mathcal{B}(\eta\to\gamma\gamma)}{\mathcal{B}(\pi^0\to\gamma\gamma)} \, \frac{\left(N_{\rm sig}^i/\epsilon_{\rm sig}^i\right)_{\Lcppi}}{\left(N_{\rm sig}^i/\epsilon_{\rm sig}^i\right)_{\Lcpeta}},
    \label{eq:bf}
\end{equation}
where $i$ denotes the c.m.~energy, $N^{i}_{\rm sig}$ is the signal yield obtained from the fit, $\epsilon^{i}_{\rm sig}$ is the signal efficiency estimated with MC simulation, $\mathcal{B}_{\pi^0\to\gamma\gamma}$ and $\mathcal{B}_{\eta\to\gamma\gamma}$ are the BFs of $\pi^0\to\gamma\gamma$ and $\eta\to\gamma\gamma$ decays taken from the PDG~\cite{ParticleDataGroup:2022pth}, respectively. The values of these variables in each fit are summarized in Table~\ref{tab:fit}.

\begin{table}[ht]
    \begin{center}
        \caption{The signal efficiencies and signal yields for each c.m.~energy. The uncertainties are statistical only.}
        \label{tab:fit}
        \small
        \begin{tabular}{@{\extracolsep{4pt}}ccccc@{}}
            \hline\hline
            \multirow{2}{*}{$\sqrt{s}\ (\rm GeV)$} & \multicolumn{2}{c}{$\Lcppi$}  & \multicolumn{2}{c}{$\Lcpeta$}                                                     \\\cline{2-3} \cline{4-5}
                                                   & $\epsilon_{\rm{sig}}^{i}(\%)$ & $N_{\rm sig}^{i}$             & $\epsilon_{\rm{sig}}^{i}(\%)$ & $N_{\rm sig}^{i}$ \\
            \hline
            4.5995                                 & $29.5\pm0.1$                  & $6.9\pm2.2$                   & $28.5\pm0.1$                  & $21.7\pm5.0$      \\
            4.6119                                 & $27.6\pm0.2$                  & $1.5\pm0.8$                   & $26.5\pm0.2$                  & $4.6\pm2.3$       \\
            4.6280                                 & $27.6\pm0.1$                  & $8.0\pm2.4$                   & $27.1\pm0.1$                  & $25.5\pm5.3$      \\
            4.6409                                 & $28.1\pm0.1$                  & $9.5\pm2.8$                   & $27.5\pm0.1$                  & $30.3\pm6.1$      \\
            4.6612                                 & $29.1\pm0.1$                  & $10.0\pm3.0$                  & $27.7\pm0.1$                  & $31.1\pm6.3$      \\
            4.6819                                 & $29.5\pm0.1$                  & $34.6\pm8.7$                  & $27.5\pm0.1$                  & $105.0\pm12.4$    \\
            4.6988                                 & $28.6\pm0.1$                  & $9.5\pm3.0$                   & $26.8\pm0.1$                  & $28.9\pm6.4$      \\
            \hline
            Total                                  & -                             & $79.9\pm18.5$                 & -                             & $247.2\pm18.1$    \\
            \hline\hline
        \end{tabular}
    \end{center}
\end{table}

The aforementioned simultaneous fits give $\mathcal{B}(\Lcppi)/\allowbreak\mathcal{B}(\Lcpeta)=0.120\pm0.026$, corresponding to integrated signal yields of $79.9\pm18.5$ for $\Lcppi$ and $247.2\pm18.1$ for $\Lcpeta$, respectively. The statistical significance of $\Lcppi$ is determined to be $5.5\sigma$, based on the change in likelihood and degrees of freedom in the fits with and without the signal. Figure~\ref{fig:fit} shows the fit projections combining all c.m.~energies.

\begin{figure}[htbp]
    \centering
    \includegraphics[width=0.4\textwidth]{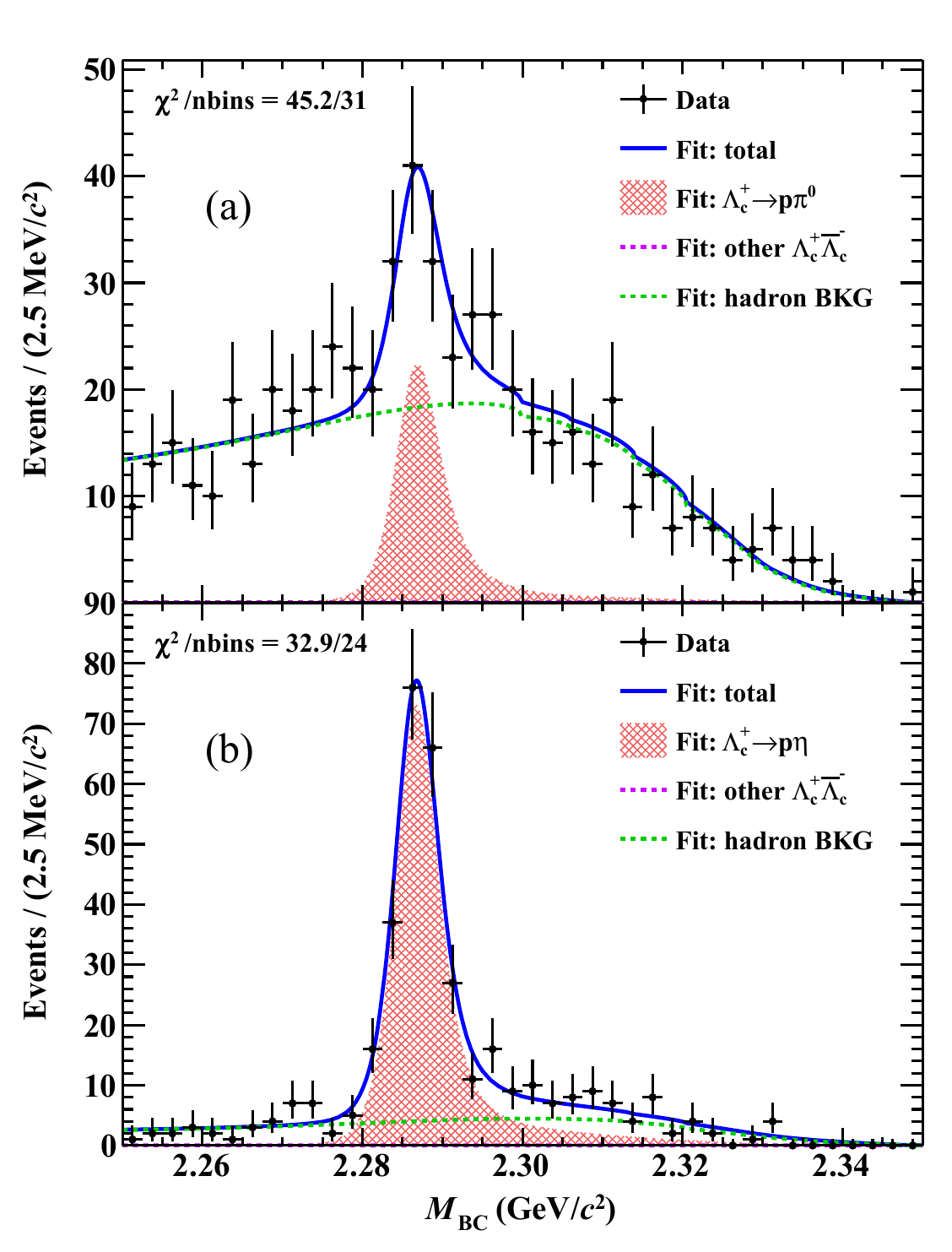}
    \caption{Post-fit $\mbc$ spectra for (a) $\Lcppi$ and (b) $\Lcpeta$ combining all c.m.~energies after implementing deep learning. The $\chi^2/\rm nbin$ values are calculated using the adaptive chi-square tests.}
    \label{fig:fit}
\end{figure}


Systematic uncertainties in the relative BF measurement arise from various sources, including $\pi^0(\eta)$ reconstruction, MC modeling, $\pi^0$ and $\eta$ decay BFs, the DNN selection, and signal yield fits. Some other uncertainties such as luminosity, cross sections, and proton and photon reconstruction cancel in the final ratio. Control samples of $J/\psi\to\pi^+\pi^-\pi^0(\eta)$~\cite{BESIII:2022ahw} ascertain a $\pi^0(\eta)$ reconstruction uncertainty of 0.5\% for $\pi^0$ and 1.0\% for $\eta$. The MC modeling uncertainty is assessed by varying the $\Lambda$ decay polarization parameter within its physical limits in the joint decay asymmetry, with the largest signal efficiency deviation found to be 0.5\% for $\Lcppi$ and 0.4\% for $\Lcpeta$. The uncertainties due to BFs of intermediate states in Eq.~\ref{eq:bf} are taken from Ref.~\cite{ParticleDataGroup:2022pth} as 0.1\% for $\pi^0\to\gamma\gamma$ and 0.5\% for $\eta\to\gamma\gamma$. A bootstrap re-sampling method~\cite{chernick2011bootstrap,BESIII:2020kzc} estimates fit-related uncertainty as 3.2\%. Here, the signal shape uncertainty is evaluated by varying the parameters of the Gaussian function. The hadron background shape uncertainty is assessed using alternative shapes, including the MC-simulated shape and another data-driven shape from c.m.~energies below the $\Lc\ALc$ production threshold at $4.47\gev$ and $4.53\gev$~\cite{BESIII:2015qfd}. The $e^+e^-\to\Lc\ALc$ background contribution is negligible.

Special considerations are given to systematic uncertainties related to the requirement on the DNN output~\cite{Shanahan:2022ifi}. As suggested by some frontier surveys~\cite{ParticleDataGroup:2022pth,Chen:2022pzc}, we consider two primary uncertainty sources, namely the {\it model uncertainty} and the {\it domain shift}. {\it Model uncertainty} is attributed to our lack of knowledge about the best model, which is estimated by altering ParT with another architecture called {\sc ParticleNet}~\cite{Qu:2019gqs} in our DNN. The {\sc ParticleNet} shares the same {\it particle cloud} representation as ParT, but processes it with a Dynamic Graph Convolutional Neural Network~\cite{Wang:2018nkf}. The relative shift in the resulting BFs, 1.9\%, is taken as the corresponding systematic uncertainty. On the other hand, {\it domain shift} describes the mismatch between datasets for training and inference, reflecting the potential data-MC inconsistency in this study. We assume such uncertainty to be mostly canceled in the relative BF due to the similarity of final states in $\Lcppi$ and $\Lcpeta$. This assumption is validated using control samples of $\Lambda^+_c\to pK^0_S\pi^0$ and $\Lambda^+_c\to pK^0_S\eta$, whose relative BF is found to be stable under any selection on their DNN outputs. The greatest deviation in the relative BF from its nominal value before implementing DNN, 4.6\%, is assigned as a residual uncertainty. In total, the systematic uncertainty for $\mathcal{B}(\Lcppi)/\mathcal{B}(\Lcpeta)$ is determined to be 6.0\% by adding the above sources in quadrature. With fit-related uncertainties considered, the final statistical significance for $\Lcppi$ is conservatively estimated to be $5.4\sigma$.


In summary, based on the data samples corresponding to an integrated luminosity of 4.5~$\ifb$ taken at c.m.~energies between 4.600 GeV and 4.699 GeV with the BESIII detector, we observe the SCS decay $\Lcppi$ for the first time with a statistical significance of $5.4\sigma$, benefiting from an innovative deep learning approach. The ratio of BFs between $\Lcppi$ and $\Lcpeta$ is measured to be $0.120\pm0.026_{\rm stat.}\pm0.007_{\rm syst.}$. Using the averaged BF for $\Lcpeta$ from BESIII~\cite{BESIII:2023ooh} and Belle~\cite{Belle:2021mvw} as reference, the BF for $\Lcppi$ is further calculated as $(1.79\pm0.39_{\rm stat.}\pm0.11_{\rm syst.}\pm0.08_{\rm ref.})\times10^{-4}$. This result agrees with the previous BESIII measurements~\cite{BESIII:2017fim,BESIII:2023uvs} and exceeds the upper limit set by Belle~\cite{Belle:2021mvw}. Figure~\ref{fig:theo} compares our result with various theoretical predictions~\cite{Uppal:1994pt,Sharma:1996sc,Chen:2002jr,Lu:2016ogy,Geng:2018plk,Cheng:2018hwl,Geng:2018rse,Geng:2019xbo,Zou:2019kzq,Zhao:2018mov,Hsiao:2021nsc,Xing:2023dni}.  This measurement supersedes those reported in Refs.~\cite{BESIII:2017fim,BESIII:2023uvs}.

The success of this study is primarily attributed to an innovative deep learning approach, adept at distinguishing signals amidst substantial backgrounds. Various reference channels and control samples at BESIII facilitate the validation and corresponding systematic uncertainty quantification of this approach. Our findings underline the great potential and broad applicability of deep learning methodologies in the context of electron-positron collider experiments.

The authors thank Huilin~Qu, Congqiao~Li and Sitian~Qian for suggestions on deep learning. The BESIII Collaboration thanks the staff of BEPCII and the IHEP computing center for their strong support. This work is supported in part by National Key R\&D Program of China under Contracts Nos. 2020YFA0406300, 2020YFA0406400, 2023YFA1606000; National Natural Science Foundation of China (NSFC) under Contracts Nos. 11635010, 11735014, 11935015, 11935016, 11935018, 12025502, 12035009, 12035013, 12061131003, 12192260, 12192261, 12192262, 12192263, 12192264, 12192265, 12221005, 12225509, 12235017, 12361141819; the Chinese Academy of Sciences (CAS) Large-Scale Scientific Facility Program; the CAS Center for Excellence in Particle Physics (CCEPP); Joint Large-Scale Scientific Facility Funds of the NSFC and CAS under Contract No. U1832207; 100 Talents Program of CAS; CAS Project for Young Scientists in Basic Research No. YSBR-117; The Institute of Nuclear and Particle Physics (INPAC) and Shanghai Key Laboratory for Particle Physics and Cosmology; German Research Foundation DFG under Contracts Nos. 455635585, FOR5327, GRK 2149; Istituto Nazionale di Fisica Nucleare, Italy; Ministry of Development of Turkey under Contract No. DPT2006K-120470; National Research Foundation of Korea under Contract No. NRF-2022R1A2C1092335; National Science and Technology fund of Mongolia; National Science Research and Innovation Fund (NSRF) via the Program Management Unit for Human Resources \& Institutional Development, Research and Innovation of Thailand under Contract No. B16F640076; Polish National Science Centre under Contract No. 2019/35/O/ST2/02907; The Swedish Research Council; U. S. Department of Energy under Contract No. DE-FG02-05ER41374

\begin{figure}[htbp]
    \centering
    \includegraphics[width=0.49\textwidth]{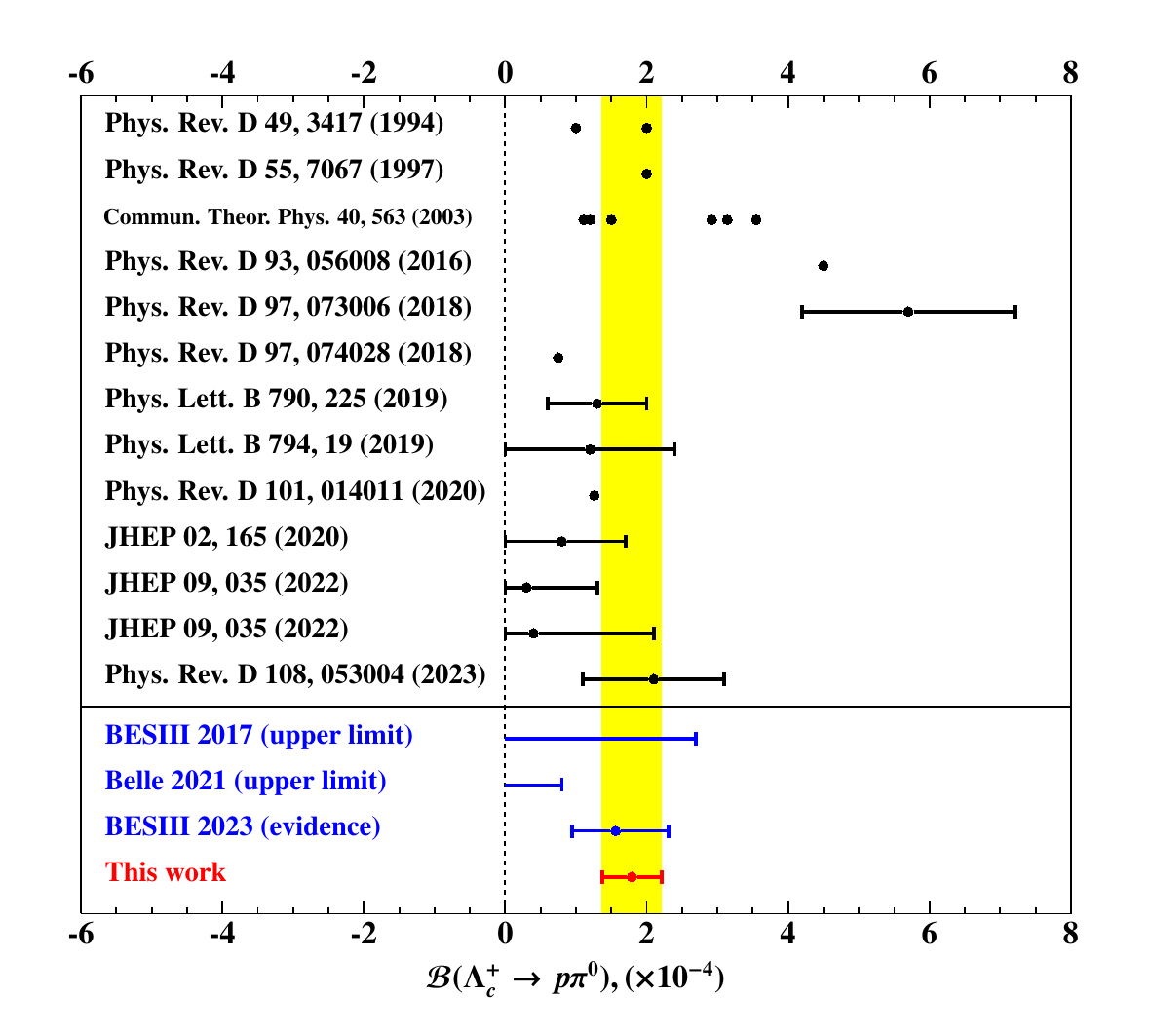}
    \caption{Comparison of our $\mathcal{B}(\Lcppi)$ result (red) with previous theoretical predictions (black) and experimental measurements (blue).}
    \label{fig:theo}
\end{figure}

\bibliographystyle{apsrev4-2}
\bibliography{main}

\end{document}

%% file: authorlist_2024-01-27.tex
\author{
\begin{small}
\begin{center}
M.~Ablikim$^{1}$, M.~N.~Achasov$^{4,c}$, P.~Adlarson$^{76}$, O.~Afedulidis$^{3}$, X.~C.~Ai$^{81}$, R.~Aliberti$^{35}$, A.~Amoroso$^{75A,75C}$, Q.~An$^{72,58,a}$, Y.~Bai$^{57}$, O.~Bakina$^{36}$
% ...
\\
\vspace{0.2cm}
(BESIII Collaboration)\\
\vspace{0.2cm} {\it
$^{1}$ Institute of High Energy Physics, Beijing 100049, People's Republic of China\\
% ...
$^{3}$ Bochum  Ruhr-University, D-44780 Bochum, Germany\\
$^{4}$ Budker Institute of Nuclear Physics SB RAS (BINP), Novosibirsk 630090, Russia\\
% ...
$^{35}$ Johannes Gutenberg University of Mainz, Johann-Joachim-Becher-Weg 45, D-55099 Mainz, Germany\\
$^{36}$ Joint Institute for Nuclear Research, 141980 Dubna, Moscow region, Russia\\
% ...
$^{57}$ Southeast University, Nanjing 211100, People's Republic of China\\
$^{58}$ State Key Laboratory of Particle Detection and Electronics, Beijing 100049, Hefei 230026, People's Republic of China\\
% ...
$^{72}$ University of Science and Technology of China, Hefei 230026, People's Republic of China\\
% ...
$^{75}$ University of Turin and INFN, (A)University of Turin, I-10125, Turin, Italy; (B)University of Eastern Piedmont, I-15121, Alessandria, Italy; (C)INFN, I-10125, Turin, Italy\\
$^{76}$ Uppsala University, Box 516, SE-75120 Uppsala, Sweden\\
% ...
$^{81}$ Zhengzhou University, Zhengzhou 450001, People's Republic of China\\
\vspace{0.2cm}
$^{a}$ Deceased\\
% ...
$^{c}$ Also at the Novosibirsk State University, Novosibirsk, 630090, Russia\\
% ...
}
\end{center}
\end{small}
\vspace{0.6cm}
}

%% file: main.bbl
\begin{thebibliography}{59}%
\makeatletter
\providecommand \@ifxundefined [1]{%
 \@ifx{#1\undefined}
}%
\providecommand \@ifnum [1]{%
 \ifnum #1\expandafter \@firstoftwo
 \else \expandafter \@secondoftwo
 \fi
}%
\providecommand \@ifx [1]{%
 \ifx #1\expandafter \@firstoftwo
 \else \expandafter \@secondoftwo
 \fi
}%
\providecommand \natexlab [1]{#1}%
\providecommand \enquote  [1]{``#1''}%
\providecommand \bibnamefont  [1]{#1}%
\providecommand \bibfnamefont [1]{#1}%
\providecommand \citenamefont [1]{#1}%
\providecommand \href@noop [0]{\@secondoftwo}%
\providecommand \href [0]{\begingroup \@sanitize@url \@href}%
\providecommand \@href[1]{\@@startlink{#1}\@@href}%
\providecommand \@@href[1]{\endgroup#1\@@endlink}%
\providecommand \@sanitize@url [0]{\catcode `\\12\catcode `\$12\catcode `\&12\catcode `\#12\catcode `\^12\catcode `\_12\catcode `\%12\relax}%
\providecommand \@@startlink[1]{}%
\providecommand \@@endlink[0]{}%
\providecommand \url  [0]{\begingroup\@sanitize@url \@url }%
\providecommand \@url [1]{\endgroup\@href {#1}{\urlprefix }}%
\providecommand \urlprefix  [0]{URL }%
\providecommand \Eprint [0]{\href }%
\providecommand \doibase [0]{https://doi.org/}%
\providecommand \selectlanguage [0]{\@gobble}%
\providecommand \bibinfo  [0]{\@secondoftwo}%
\providecommand \bibfield  [0]{\@secondoftwo}%
\providecommand \translation [1]{[#1]}%
\providecommand \BibitemOpen [0]{}%
\providecommand \bibitemStop [0]{}%
\providecommand \bibitemNoStop [0]{.\EOS\space}%
\providecommand \EOS [0]{\spacefactor3000\relax}%
\providecommand \BibitemShut  [1]{\csname bibitem#1\endcsname}%
\let\auto@bib@innerbib\@empty
\bibitem [{\citenamefont {Cheng}(2022)}]{Cheng:2021qpd}%
  \BibitemOpen
  \bibfield  {author} {\bibinfo {author} {\bibfnamefont {H.~Y.}\ \bibnamefont {Cheng}},\ }\href {https://doi.org/10.1016/j.cjph.2022.06.021} {\bibfield  {journal} {\bibinfo  {journal} {Chin. J. Phys.}\ }\textbf {\bibinfo {volume} {78}},\ \bibinfo {pages} {324} (\bibinfo {year} {2022})},\ \Eprint {https://arxiv.org/abs/2109.01216} {arXiv:2109.01216 [hep-ph]} \BibitemShut {NoStop}%
\bibitem [{\citenamefont {Chau}(1983)}]{Chau:1982da}%
  \BibitemOpen
  \bibfield  {author} {\bibinfo {author} {\bibfnamefont {L.~L.}\ \bibnamefont {Chau}},\ }\href {https://doi.org/10.1016/0370-1573(83)90043-1} {\bibfield  {journal} {\bibinfo  {journal} {Phys. Rept.}\ }\textbf {\bibinfo {volume} {95}},\ \bibinfo {pages} {1} (\bibinfo {year} {1983})}\BibitemShut {NoStop}%
\bibitem [{\citenamefont {Chau}\ and\ \citenamefont {Cheng}(1986)}]{Chau:1986jb}%
  \BibitemOpen
  \bibfield  {author} {\bibinfo {author} {\bibfnamefont {L.~L.}\ \bibnamefont {Chau}}\ and\ \bibinfo {author} {\bibfnamefont {H.~Y.}\ \bibnamefont {Cheng}},\ }\href {https://doi.org/10.1103/PhysRevLett.56.1655} {\bibfield  {journal} {\bibinfo  {journal} {Phys. Rev. Lett.}\ }\textbf {\bibinfo {volume} {56}},\ \bibinfo {pages} {1655} (\bibinfo {year} {1986})}\BibitemShut {NoStop}%
\bibitem [{\citenamefont {Chau}\ \emph {et~al.}(1996)\citenamefont {Chau}, \citenamefont {Cheng},\ and\ \citenamefont {Tseng}}]{Chau:1995gk}%
  \BibitemOpen
  \bibfield  {author} {\bibinfo {author} {\bibfnamefont {L.~L.}\ \bibnamefont {Chau}}, \bibinfo {author} {\bibfnamefont {H.~Y.}\ \bibnamefont {Cheng}},\ and\ \bibinfo {author} {\bibfnamefont {B.}~\bibnamefont {Tseng}},\ }\href {https://doi.org/10.1103/PhysRevD.54.2132} {\bibfield  {journal} {\bibinfo  {journal} {Phys. Rev. D}\ }\textbf {\bibinfo {volume} {54}},\ \bibinfo {pages} {2132} (\bibinfo {year} {1996})},\ \Eprint {https://arxiv.org/abs/hep-ph/9508382} {arXiv:hep-ph/9508382} \BibitemShut {NoStop}%
\bibitem [{\citenamefont {Cheng}\ and\ \citenamefont {Tseng}(1993)}]{Cheng:1993gf}%
  \BibitemOpen
  \bibfield  {author} {\bibinfo {author} {\bibfnamefont {H.~Y.}\ \bibnamefont {Cheng}}\ and\ \bibinfo {author} {\bibfnamefont {B.}~\bibnamefont {Tseng}},\ }\href {https://doi.org/10.1103/PhysRevD.48.4188} {\bibfield  {journal} {\bibinfo  {journal} {Phys. Rev. D}\ }\textbf {\bibinfo {volume} {48}},\ \bibinfo {pages} {4188} (\bibinfo {year} {1993})},\ \Eprint {https://arxiv.org/abs/hep-ph/9304286} {arXiv:hep-ph/9304286} \BibitemShut {NoStop}%
\bibitem [{\citenamefont {Kohara}(1998)}]{Kohara:1998jm}%
  \BibitemOpen
  \bibfield  {author} {\bibinfo {author} {\bibfnamefont {Y.}~\bibnamefont {Kohara}},\ }\href@noop {} {\bibfield  {journal} {\bibinfo  {journal} {Nuovo Cim. A}\ }\textbf {\bibinfo {volume} {111}},\ \bibinfo {pages} {67} (\bibinfo {year} {1998})}\BibitemShut {NoStop}%
\bibitem [{\citenamefont {Ke}\ \emph {et~al.}(2023)\citenamefont {Ke}, \citenamefont {Koponen}, \citenamefont {Li},\ and\ \citenamefont {Zheng}}]{Ke:2023qzc}%
  \BibitemOpen
  \bibfield  {author} {\bibinfo {author} {\bibfnamefont {B.~C.}\ \bibnamefont {Ke}}, \bibinfo {author} {\bibfnamefont {J.}~\bibnamefont {Koponen}}, \bibinfo {author} {\bibfnamefont {H.~B.}\ \bibnamefont {Li}},\ and\ \bibinfo {author} {\bibfnamefont {Y.~H.}\ \bibnamefont {Zheng}},\ }\href {https://doi.org/10.1146/annurev-nucl-110222-044046} {\bibfield  {journal} {\bibinfo  {journal} {Ann. Rev. Nucl. Part. Sci.}\ }\textbf {\bibinfo {volume} {73}},\ \bibinfo {pages} {285} (\bibinfo {year} {2023})},\ \Eprint {https://arxiv.org/abs/2310.05228} {arXiv:2310.05228 [hep-ex]} \BibitemShut {NoStop}%
\bibitem [{\citenamefont {Li}\ and\ \citenamefont {Lyu}(2021)}]{Li:2021iwf}%
  \BibitemOpen
  \bibfield  {author} {\bibinfo {author} {\bibfnamefont {H.~B.}\ \bibnamefont {Li}}\ and\ \bibinfo {author} {\bibfnamefont {X.~R.}\ \bibnamefont {Lyu}},\ }\href {https://doi.org/10.1093/nsr/nwab181} {\bibfield  {journal} {\bibinfo  {journal} {Natl. Sci. Rev.}\ }\textbf {\bibinfo {volume} {8}},\ \bibinfo {pages} {nwab181} (\bibinfo {year} {2021})},\ \Eprint {https://arxiv.org/abs/2103.00908} {arXiv:2103.00908 [hep-ex]} \BibitemShut {NoStop}%
\bibitem [{\citenamefont {Uppal}\ \emph {et~al.}(1994)\citenamefont {Uppal}, \citenamefont {Verma},\ and\ \citenamefont {Khanna}}]{Uppal:1994pt}%
  \BibitemOpen
  \bibfield  {author} {\bibinfo {author} {\bibfnamefont {T.}~\bibnamefont {Uppal}}, \bibinfo {author} {\bibfnamefont {R.~C.}\ \bibnamefont {Verma}},\ and\ \bibinfo {author} {\bibfnamefont {M.~P.}\ \bibnamefont {Khanna}},\ }\href {https://doi.org/10.1103/PhysRevD.49.3417} {\bibfield  {journal} {\bibinfo  {journal} {Phys. Rev. D}\ }\textbf {\bibinfo {volume} {49}},\ \bibinfo {pages} {3417} (\bibinfo {year} {1994})}\BibitemShut {NoStop}%
\bibitem [{\citenamefont {Chen}\ \emph {et~al.}(2003)\citenamefont {Chen}, \citenamefont {Guo}, \citenamefont {Li},\ and\ \citenamefont {Wang}}]{Chen:2002jr}%
  \BibitemOpen
  \bibfield  {author} {\bibinfo {author} {\bibfnamefont {S.~L.}\ \bibnamefont {Chen}}, \bibinfo {author} {\bibfnamefont {X.~H.}\ \bibnamefont {Guo}}, \bibinfo {author} {\bibfnamefont {X.~Q.}\ \bibnamefont {Li}},\ and\ \bibinfo {author} {\bibfnamefont {G.~L.}\ \bibnamefont {Wang}},\ }\href {https://doi.org/10.1088/0253-6102/40/5/563} {\bibfield  {journal} {\bibinfo  {journal} {Commun. Theor. Phys.}\ }\textbf {\bibinfo {volume} {40}},\ \bibinfo {pages} {563} (\bibinfo {year} {2003})},\ \Eprint {https://arxiv.org/abs/hep-ph/0208006} {arXiv:hep-ph/0208006} \BibitemShut {NoStop}%
\bibitem [{\citenamefont {Cheng}\ \emph {et~al.}(2018)\citenamefont {Cheng}, \citenamefont {Kang},\ and\ \citenamefont {Xu}}]{Cheng:2018hwl}%
  \BibitemOpen
  \bibfield  {author} {\bibinfo {author} {\bibfnamefont {H.~Y.}\ \bibnamefont {Cheng}}, \bibinfo {author} {\bibfnamefont {X.~W.}\ \bibnamefont {Kang}},\ and\ \bibinfo {author} {\bibfnamefont {F.}~\bibnamefont {Xu}},\ }\href {https://doi.org/10.1103/PhysRevD.97.074028} {\bibfield  {journal} {\bibinfo  {journal} {Phys. Rev. D}\ }\textbf {\bibinfo {volume} {97}},\ \bibinfo {pages} {074028} (\bibinfo {year} {2018})},\ \Eprint {https://arxiv.org/abs/1801.08625} {arXiv:1801.08625 [hep-ph]} \BibitemShut {NoStop}%
\bibitem [{\citenamefont {Zou}\ \emph {et~al.}(2020)\citenamefont {Zou}, \citenamefont {Xu}, \citenamefont {Meng},\ and\ \citenamefont {Cheng}}]{Zou:2019kzq}%
  \BibitemOpen
  \bibfield  {author} {\bibinfo {author} {\bibfnamefont {J.}~\bibnamefont {Zou}}, \bibinfo {author} {\bibfnamefont {F.}~\bibnamefont {Xu}}, \bibinfo {author} {\bibfnamefont {G.}~\bibnamefont {Meng}},\ and\ \bibinfo {author} {\bibfnamefont {H.~Y.}\ \bibnamefont {Cheng}},\ }\href {https://doi.org/10.1103/PhysRevD.101.014011} {\bibfield  {journal} {\bibinfo  {journal} {Phys. Rev. D}\ }\textbf {\bibinfo {volume} {101}},\ \bibinfo {pages} {014011} (\bibinfo {year} {2020})},\ \Eprint {https://arxiv.org/abs/1910.13626} {arXiv:1910.13626 [hep-ph]} \BibitemShut {NoStop}%
\bibitem [{\citenamefont {Zhao}\ \emph {et~al.}(2020)\citenamefont {Zhao}, \citenamefont {Wang}, \citenamefont {Hsiao},\ and\ \citenamefont {Yu}}]{Zhao:2018mov}%
  \BibitemOpen
  \bibfield  {author} {\bibinfo {author} {\bibfnamefont {H.~J.}\ \bibnamefont {Zhao}}, \bibinfo {author} {\bibfnamefont {Y.~L.}\ \bibnamefont {Wang}}, \bibinfo {author} {\bibfnamefont {Y.~K.}\ \bibnamefont {Hsiao}},\ and\ \bibinfo {author} {\bibfnamefont {Y.}~\bibnamefont {Yu}},\ }\href {https://doi.org/10.1007/JHEP02(2020)165} {\bibfield  {journal} {\bibinfo  {journal} {JHEP}\ }\textbf {\bibinfo {volume} {02}},\ \bibinfo {pages} {165 (2020)}},\ \Eprint {https://arxiv.org/abs/1811.07265} {arXiv:1811.07265 [hep-ph]} \BibitemShut {NoStop}%
\bibitem [{\citenamefont {Hsiao}\ \emph {et~al.}(2022)\citenamefont {Hsiao}, \citenamefont {Wang},\ and\ \citenamefont {Zhao}}]{Hsiao:2021nsc}%
  \BibitemOpen
  \bibfield  {author} {\bibinfo {author} {\bibfnamefont {Y.~K.}\ \bibnamefont {Hsiao}}, \bibinfo {author} {\bibfnamefont {Y.~L.}\ \bibnamefont {Wang}},\ and\ \bibinfo {author} {\bibfnamefont {H.~J.}\ \bibnamefont {Zhao}},\ }\href {https://doi.org/10.1007/JHEP09(2022)035} {\bibfield  {journal} {\bibinfo  {journal} {JHEP}\ }\textbf {\bibinfo {volume} {09}},\ \bibinfo {pages} {035 (2022)}},\ \Eprint {https://arxiv.org/abs/2111.04124} {arXiv:2111.04124 [hep-ph]} \BibitemShut {NoStop}%
\bibitem [{\citenamefont {Sharma}\ and\ \citenamefont {Verma}(1997)}]{Sharma:1996sc}%
  \BibitemOpen
  \bibfield  {author} {\bibinfo {author} {\bibfnamefont {K.~K.}\ \bibnamefont {Sharma}}\ and\ \bibinfo {author} {\bibfnamefont {R.~C.}\ \bibnamefont {Verma}},\ }\href {https://doi.org/10.1103/PhysRevD.55.7067} {\bibfield  {journal} {\bibinfo  {journal} {Phys. Rev. D}\ }\textbf {\bibinfo {volume} {55}},\ \bibinfo {pages} {7067} (\bibinfo {year} {1997})},\ \Eprint {https://arxiv.org/abs/hep-ph/9704391} {arXiv:hep-ph/9704391} \BibitemShut {NoStop}%
\bibitem [{\citenamefont {L\"u}\ \emph {et~al.}(2016)\citenamefont {L\"u}, \citenamefont {Wang},\ and\ \citenamefont {Yu}}]{Lu:2016ogy}%
  \BibitemOpen
  \bibfield  {author} {\bibinfo {author} {\bibfnamefont {C.~D.}\ \bibnamefont {L\"u}}, \bibinfo {author} {\bibfnamefont {W.}~\bibnamefont {Wang}},\ and\ \bibinfo {author} {\bibfnamefont {F.~S.}\ \bibnamefont {Yu}},\ }\href {https://doi.org/10.1103/PhysRevD.93.056008} {\bibfield  {journal} {\bibinfo  {journal} {Phys. Rev. D}\ }\textbf {\bibinfo {volume} {93}},\ \bibinfo {pages} {056008} (\bibinfo {year} {2016})},\ \Eprint {https://arxiv.org/abs/1601.04241} {arXiv:1601.04241 [hep-ph]} \BibitemShut {NoStop}%
\bibitem [{\citenamefont {Geng}\ \emph {et~al.}(2018)\citenamefont {Geng}, \citenamefont {Hsiao}, \citenamefont {Liu},\ and\ \citenamefont {Tsai}}]{Geng:2018plk}%
  \BibitemOpen
  \bibfield  {author} {\bibinfo {author} {\bibfnamefont {C.~Q.}\ \bibnamefont {Geng}}, \bibinfo {author} {\bibfnamefont {Y.~K.}\ \bibnamefont {Hsiao}}, \bibinfo {author} {\bibfnamefont {C.~W.}\ \bibnamefont {Liu}},\ and\ \bibinfo {author} {\bibfnamefont {T.~H.}\ \bibnamefont {Tsai}},\ }\href {https://doi.org/10.1103/PhysRevD.97.073006} {\bibfield  {journal} {\bibinfo  {journal} {Phys. Rev. D}\ }\textbf {\bibinfo {volume} {97}},\ \bibinfo {pages} {073006} (\bibinfo {year} {2018})},\ \Eprint {https://arxiv.org/abs/1801.03276} {arXiv:1801.03276 [hep-ph]} \BibitemShut {NoStop}%
\bibitem [{\citenamefont {Geng}\ \emph {et~al.}(2019{\natexlab{a}})\citenamefont {Geng}, \citenamefont {Liu},\ and\ \citenamefont {Tsai}}]{Geng:2018rse}%
  \BibitemOpen
  \bibfield  {author} {\bibinfo {author} {\bibfnamefont {C.~Q.}\ \bibnamefont {Geng}}, \bibinfo {author} {\bibfnamefont {C.~W.}\ \bibnamefont {Liu}},\ and\ \bibinfo {author} {\bibfnamefont {T.~H.}\ \bibnamefont {Tsai}},\ }\href {https://doi.org/10.1016/j.physletb.2019.01.025} {\bibfield  {journal} {\bibinfo  {journal} {Phys. Lett. B}\ }\textbf {\bibinfo {volume} {790}},\ \bibinfo {pages} {225} (\bibinfo {year} {2019}{\natexlab{a}})},\ \Eprint {https://arxiv.org/abs/1812.08508} {arXiv:1812.08508 [hep-ph]} \BibitemShut {NoStop}%
\bibitem [{\citenamefont {Geng}\ \emph {et~al.}(2019{\natexlab{b}})\citenamefont {Geng}, \citenamefont {Liu},\ and\ \citenamefont {Tsai}}]{Geng:2019xbo}%
  \BibitemOpen
  \bibfield  {author} {\bibinfo {author} {\bibfnamefont {C.~Q.}\ \bibnamefont {Geng}}, \bibinfo {author} {\bibfnamefont {C.~W.}\ \bibnamefont {Liu}},\ and\ \bibinfo {author} {\bibfnamefont {T.~H.}\ \bibnamefont {Tsai}},\ }\href {https://doi.org/10.1016/j.physletb.2019.05.024} {\bibfield  {journal} {\bibinfo  {journal} {Phys. Lett. B}\ }\textbf {\bibinfo {volume} {794}},\ \bibinfo {pages} {19} (\bibinfo {year} {2019}{\natexlab{b}})},\ \Eprint {https://arxiv.org/abs/1902.06189} {arXiv:1902.06189 [hep-ph]} \BibitemShut {NoStop}%
\bibitem [{\citenamefont {Xing}\ \emph {et~al.}(2023)\citenamefont {Xing}, \citenamefont {He}, \citenamefont {Huang},\ and\ \citenamefont {Yang}}]{Xing:2023dni}%
  \BibitemOpen
  \bibfield  {author} {\bibinfo {author} {\bibfnamefont {Z.~P.}\ \bibnamefont {Xing}}, \bibinfo {author} {\bibfnamefont {X.~G.}\ \bibnamefont {He}}, \bibinfo {author} {\bibfnamefont {F.}~\bibnamefont {Huang}},\ and\ \bibinfo {author} {\bibfnamefont {C.}~\bibnamefont {Yang}},\ }\href {https://doi.org/10.1103/PhysRevD.108.053004} {\bibfield  {journal} {\bibinfo  {journal} {Phys. Rev. D}\ }\textbf {\bibinfo {volume} {108}},\ \bibinfo {pages} {053004} (\bibinfo {year} {2023})}\BibitemShut {NoStop}%
\bibitem [{\citenamefont {Ablikim}\ \emph {et~al.}(2024)\citenamefont {Ablikim} \emph {et~al.}}]{BESIII:2023uvs}%
  \BibitemOpen
  \bibfield  {author} {\bibinfo {author} {\bibfnamefont {M.}~\bibnamefont {Ablikim}} \emph {et~al.} (\bibinfo {collaboration} {BESIII}),\ }\href {https://doi.org/10.1103/PhysRevD.109.L091101} {\bibfield  {journal} {\bibinfo  {journal} {Phys. Rev. D}\ }\textbf {\bibinfo {volume} {109}},\ \bibinfo {pages} {L091101} (\bibinfo {year} {2024})},\ \Eprint {https://arxiv.org/abs/2311.06883} {arXiv:2311.06883 [hep-ex]} \BibitemShut {NoStop}%
\bibitem [{\citenamefont {Li}\ \emph {et~al.}(2021)\citenamefont {Li} \emph {et~al.}}]{Belle:2021mvw}%
  \BibitemOpen
  \bibfield  {author} {\bibinfo {author} {\bibfnamefont {S.~X.}\ \bibnamefont {Li}} \emph {et~al.} (\bibinfo {collaboration} {Belle Collaboration}),\ }\href {https://doi.org/10.1103/PhysRevD.103.072004} {\bibfield  {journal} {\bibinfo  {journal} {Phys. Rev. D}\ }\textbf {\bibinfo {volume} {103}},\ \bibinfo {pages} {072004} (\bibinfo {year} {2021})},\ \Eprint {https://arxiv.org/abs/2102.12226} {arXiv:2102.12226 [hep-ex]} \BibitemShut {NoStop}%
\bibitem [{\citenamefont {Baltrusaitis}\ \emph {et~al.}(1986)\citenamefont {Baltrusaitis} \emph {et~al.}}]{MARK-III:1985hbd}%
  \BibitemOpen
  \bibfield  {author} {\bibinfo {author} {\bibfnamefont {R.~M.}\ \bibnamefont {Baltrusaitis}} \emph {et~al.} (\bibinfo {collaboration} {MARK-III Collaboration}),\ }\href {https://doi.org/10.1103/PhysRevLett.56.2140} {\bibfield  {journal} {\bibinfo  {journal} {Phys. Rev. Lett.}\ }\textbf {\bibinfo {volume} {56}},\ \bibinfo {pages} {2140} (\bibinfo {year} {1986})}\BibitemShut {NoStop}%
\bibitem [{\citenamefont {LeCun}\ \emph {et~al.}(2015)\citenamefont {LeCun}, \citenamefont {Bengio},\ and\ \citenamefont {Hinton}}]{deeplearning}%
  \BibitemOpen
  \bibfield  {author} {\bibinfo {author} {\bibfnamefont {Y.}~\bibnamefont {LeCun}}, \bibinfo {author} {\bibfnamefont {Y.}~\bibnamefont {Bengio}},\ and\ \bibinfo {author} {\bibfnamefont {G.}~\bibnamefont {Hinton}},\ }\href {https://doi.org/10.1038/nature14539} {\bibfield  {journal} {\bibinfo  {journal} {Nature}\ }\textbf {\bibinfo {volume} {521}},\ \bibinfo {pages} {436} (\bibinfo {year} {2015})}\BibitemShut {NoStop}%
\bibitem [{\citenamefont {Schwartz}(2021)}]{Schwartz2021Modern}%
  \BibitemOpen
  \bibfield  {author} {\bibinfo {author} {\bibfnamefont {M.~D.}\ \bibnamefont {Schwartz}},\ }\href {https://doi.org/10.1162/99608f92.beeb1183} {\bibfield  {journal} {\bibinfo  {journal} {Harvard Data Science Review}\ }\textbf {\bibinfo {volume} {3}},\ \bibinfo {pages} {2} (\bibinfo {year} {2021})}\BibitemShut {NoStop}%
\bibitem [{\citenamefont {Larkoski}\ \emph {et~al.}(2020)\citenamefont {Larkoski}, \citenamefont {Moult},\ and\ \citenamefont {Nachman}}]{Larkoski:2017jix}%
  \BibitemOpen
  \bibfield  {author} {\bibinfo {author} {\bibfnamefont {A.~J.}\ \bibnamefont {Larkoski}}, \bibinfo {author} {\bibfnamefont {I.}~\bibnamefont {Moult}},\ and\ \bibinfo {author} {\bibfnamefont {B.}~\bibnamefont {Nachman}},\ }\href {https://doi.org/10.1016/j.physrep.2019.11.001} {\bibfield  {journal} {\bibinfo  {journal} {Phys. Rept.}\ }\textbf {\bibinfo {volume} {841}},\ \bibinfo {pages} {1} (\bibinfo {year} {2020})},\ \Eprint {https://arxiv.org/abs/1709.04464} {arXiv:1709.04464 [hep-ph]} \BibitemShut {NoStop}%
\bibitem [{\citenamefont {Kogler}\ \emph {et~al.}(2019)\citenamefont {Kogler} \emph {et~al.}}]{Kogler:2018hem}%
  \BibitemOpen
  \bibfield  {author} {\bibinfo {author} {\bibfnamefont {R.}~\bibnamefont {Kogler}} \emph {et~al.},\ }\href {https://doi.org/10.1103/RevModPhys.91.045003} {\bibfield  {journal} {\bibinfo  {journal} {Rev. Mod. Phys.}\ }\textbf {\bibinfo {volume} {91}},\ \bibinfo {pages} {045003} (\bibinfo {year} {2019})},\ \Eprint {https://arxiv.org/abs/1803.06991} {arXiv:1803.06991 [hep-ex]} \BibitemShut {NoStop}%
\bibitem [{\citenamefont {{HEP ML Community}}()}]{hepmllivingreview}%
  \BibitemOpen
  \bibfield  {author} {\bibinfo {author} {\bibnamefont {{HEP ML Community}}},\ }\href {https://iml-wg.github.io/HEPML-LivingReview/} {\bibinfo {title} {{A Living Review of Machine Learning for Particle Physics}}}\BibitemShut {NoStop}%
\bibitem [{\citenamefont {Ablikim}\ \emph {et~al.}(2022{\natexlab{a}})\citenamefont {Ablikim} \emph {et~al.}}]{BESIII:2022dxl}%
  \BibitemOpen
  \bibfield  {author} {\bibinfo {author} {\bibfnamefont {M.}~\bibnamefont {Ablikim}} \emph {et~al.} (\bibinfo {collaboration} {BESIII Collaboration}),\ }\href {https://doi.org/10.1088/1674-1137/ac80b4} {\bibfield  {journal} {\bibinfo  {journal} {Chin. Phys. C}\ }\textbf {\bibinfo {volume} {46}},\ \bibinfo {pages} {113002} (\bibinfo {year} {2022}{\natexlab{a}})},\ \Eprint {https://arxiv.org/abs/2203.03133} {arXiv:2203.03133 [hep-ex]} \BibitemShut {NoStop}%
\bibitem [{\citenamefont {Ablikim}\ \emph {et~al.}(2022{\natexlab{b}})\citenamefont {Ablikim} \emph {et~al.}}]{BESIII:2022ulv}%
  \BibitemOpen
  \bibfield  {author} {\bibinfo {author} {\bibfnamefont {M.}~\bibnamefont {Ablikim}} \emph {et~al.} (\bibinfo {collaboration} {BESIII Collaboration}),\ }\href {https://doi.org/10.1088/1674-1137/ac84cc} {\bibfield  {journal} {\bibinfo  {journal} {Chin. Phys. C}\ }\textbf {\bibinfo {volume} {46}},\ \bibinfo {pages} {113003} (\bibinfo {year} {2022}{\natexlab{b}})},\ \Eprint {https://arxiv.org/abs/2205.04809} {arXiv:2205.04809 [hep-ex]} \BibitemShut {NoStop}%
\bibitem [{\citenamefont {Ablikim}\ \emph {et~al.}(2010)\citenamefont {Ablikim} \emph {et~al.}}]{BESIII:2009fln}%
  \BibitemOpen
  \bibfield  {author} {\bibinfo {author} {\bibfnamefont {M.}~\bibnamefont {Ablikim}} \emph {et~al.} (\bibinfo {collaboration} {BESIII Collaboration}),\ }\href {https://doi.org/10.1016/j.nima.2009.12.050} {\bibfield  {journal} {\bibinfo  {journal} {Nucl. Instrum. Meth. A}\ }\textbf {\bibinfo {volume} {614}},\ \bibinfo {pages} {345} (\bibinfo {year} {2010})},\ \Eprint {https://arxiv.org/abs/0911.4960} {arXiv:0911.4960 [physics.ins-det]} \BibitemShut {NoStop}%
\bibitem [{\citenamefont {Ablikim}\ \emph {et~al.}(2022{\natexlab{c}})\citenamefont {Ablikim} \emph {et~al.}}]{BESIII:2022bkj}%
  \BibitemOpen
  \bibfield  {author} {\bibinfo {author} {\bibfnamefont {M.}~\bibnamefont {Ablikim}} \emph {et~al.} (\bibinfo {collaboration} {BESIII Collaboration}),\ }\href {https://doi.org/10.1103/PhysRevLett.128.142001} {\bibfield  {journal} {\bibinfo  {journal} {Phys. Rev. Lett.}\ }\textbf {\bibinfo {volume} {128}},\ \bibinfo {pages} {142001} (\bibinfo {year} {2022}{\natexlab{c}})},\ \Eprint {https://arxiv.org/abs/2201.02056} {arXiv:2201.02056 [hep-ex]} \BibitemShut {NoStop}%
\bibitem [{\citenamefont {Ablikim}\ \emph {et~al.}(2017)\citenamefont {Ablikim} \emph {et~al.}}]{BESIII:2017fim}%
  \BibitemOpen
  \bibfield  {author} {\bibinfo {author} {\bibfnamefont {M.}~\bibnamefont {Ablikim}} \emph {et~al.} (\bibinfo {collaboration} {BESIII Collaboration}),\ }\href {https://doi.org/10.1103/PhysRevD.95.111102} {\bibfield  {journal} {\bibinfo  {journal} {Phys. Rev. D}\ }\textbf {\bibinfo {volume} {95}},\ \bibinfo {pages} {111102} (\bibinfo {year} {2017})},\ \Eprint {https://arxiv.org/abs/1702.05279} {arXiv:1702.05279 [hep-ex]} \BibitemShut {NoStop}%
\bibitem [{\citenamefont {Ablikim}\ \emph {et~al.}(2023)\citenamefont {Ablikim} \emph {et~al.}}]{BESIII:2023ooh}%
  \BibitemOpen
  \bibfield  {author} {\bibinfo {author} {\bibfnamefont {M.}~\bibnamefont {Ablikim}} \emph {et~al.} (\bibinfo {collaboration} {BESIII Collaboration}),\ }\href {https://doi.org/10.1007/JHEP11(2023)137} {\bibfield  {journal} {\bibinfo  {journal} {JHEP}\ }\textbf {\bibinfo {volume} {11}},\ \bibinfo {pages} {137 (2023)}},\ \Eprint {https://arxiv.org/abs/2307.09266} {arXiv:2307.09266 [hep-ex]} \BibitemShut {NoStop}%
\bibitem [{\citenamefont {Voss}\ \emph {et~al.}(2009)\citenamefont {Voss}, \citenamefont {Höcker}, \citenamefont {Stelzer},\ and\ \citenamefont {Tegenfeldt}}]{Voss:2009rK}%
  \BibitemOpen
  \bibfield  {author} {\bibinfo {author} {\bibfnamefont {H.}~\bibnamefont {Voss}}, \bibinfo {author} {\bibfnamefont {A.}~\bibnamefont {Höcker}}, \bibinfo {author} {\bibfnamefont {J.}~\bibnamefont {Stelzer}},\ and\ \bibinfo {author} {\bibfnamefont {F.}~\bibnamefont {Tegenfeldt}},\ }\href {https://doi.org/10.22323/1.050.0040} {\bibfield  {journal} {\bibinfo  {journal} {PoS}\ }\textbf {\bibinfo {volume} {ACAT}},\ \bibinfo {pages} {040} (\bibinfo {year} {2009})}\BibitemShut {NoStop}%
\bibitem [{\citenamefont {de~Oliveira}\ \emph {et~al.}(2016)\citenamefont {de~Oliveira}, \citenamefont {Kagan}, \citenamefont {Mackey}, \citenamefont {Nachman},\ and\ \citenamefont {Schwartzman}}]{deOliveira:2015xxd}%
  \BibitemOpen
  \bibfield  {author} {\bibinfo {author} {\bibfnamefont {L.}~\bibnamefont {de~Oliveira}}, \bibinfo {author} {\bibfnamefont {M.}~\bibnamefont {Kagan}}, \bibinfo {author} {\bibfnamefont {L.}~\bibnamefont {Mackey}}, \bibinfo {author} {\bibfnamefont {B.}~\bibnamefont {Nachman}},\ and\ \bibinfo {author} {\bibfnamefont {A.}~\bibnamefont {Schwartzman}},\ }\href {https://doi.org/10.1007/JHEP07(2016)069} {\bibfield  {journal} {\bibinfo  {journal} {JHEP}\ }\textbf {\bibinfo {volume} {07}},\ \bibinfo {pages} {069 (2016)}},\ \Eprint {https://arxiv.org/abs/1511.05190} {arXiv:1511.05190 [hep-ph]} \BibitemShut {NoStop}%
\bibitem [{\citenamefont {Guest}\ \emph {et~al.}(2016)\citenamefont {Guest}, \citenamefont {Collado}, \citenamefont {Baldi}, \citenamefont {Hsu}, \citenamefont {Urban},\ and\ \citenamefont {Whiteson}}]{Guest:2016iqz}%
  \BibitemOpen
  \bibfield  {author} {\bibinfo {author} {\bibfnamefont {D.}~\bibnamefont {Guest}}, \bibinfo {author} {\bibfnamefont {J.}~\bibnamefont {Collado}}, \bibinfo {author} {\bibfnamefont {P.}~\bibnamefont {Baldi}}, \bibinfo {author} {\bibfnamefont {S.~C.}\ \bibnamefont {Hsu}}, \bibinfo {author} {\bibfnamefont {G.}~\bibnamefont {Urban}},\ and\ \bibinfo {author} {\bibfnamefont {D.}~\bibnamefont {Whiteson}},\ }\href {https://doi.org/10.1103/PhysRevD.94.112002} {\bibfield  {journal} {\bibinfo  {journal} {Phys. Rev. D}\ }\textbf {\bibinfo {volume} {94}},\ \bibinfo {pages} {112002} (\bibinfo {year} {2016})},\ \Eprint {https://arxiv.org/abs/1607.08633} {arXiv:1607.08633 [hep-ex]} \BibitemShut {NoStop}%
\bibitem [{\citenamefont {Louppe}\ \emph {et~al.}(2019)\citenamefont {Louppe}, \citenamefont {Cho}, \citenamefont {Becot},\ and\ \citenamefont {Cranmer}}]{Louppe:2017ipp}%
  \BibitemOpen
  \bibfield  {author} {\bibinfo {author} {\bibfnamefont {G.}~\bibnamefont {Louppe}}, \bibinfo {author} {\bibfnamefont {K.}~\bibnamefont {Cho}}, \bibinfo {author} {\bibfnamefont {C.}~\bibnamefont {Becot}},\ and\ \bibinfo {author} {\bibfnamefont {K.}~\bibnamefont {Cranmer}},\ }\href {https://doi.org/10.1007/JHEP01(2019)057} {\bibfield  {journal} {\bibinfo  {journal} {JHEP}\ }\textbf {\bibinfo {volume} {01}},\ \bibinfo {pages} {057 (2019)}},\ \Eprint {https://arxiv.org/abs/1702.00748} {arXiv:1702.00748 [hep-ph]} \BibitemShut {NoStop}%
\bibitem [{\citenamefont {Henrion}\ \emph {et~al.}(2017)\citenamefont {Henrion}, \citenamefont {Brehmer}, \citenamefont {Bruna}, \citenamefont {Cho}, \citenamefont {Cranmer}, \citenamefont {Louppe},\ and\ \citenamefont {Rochette}}]{henrion2017neural}%
  \BibitemOpen
  \bibfield  {author} {\bibinfo {author} {\bibfnamefont {I.}~\bibnamefont {Henrion}}, \bibinfo {author} {\bibfnamefont {J.}~\bibnamefont {Brehmer}}, \bibinfo {author} {\bibfnamefont {J.}~\bibnamefont {Bruna}}, \bibinfo {author} {\bibfnamefont {K.}~\bibnamefont {Cho}}, \bibinfo {author} {\bibfnamefont {K.}~\bibnamefont {Cranmer}}, \bibinfo {author} {\bibfnamefont {G.}~\bibnamefont {Louppe}},\ and\ \bibinfo {author} {\bibfnamefont {G.}~\bibnamefont {Rochette}},\ }in\ \href@noop {} {\emph {\bibinfo {booktitle} {Deep Learning for Physical Sciences Workshop}}}\ (\bibinfo {organization} {NIPS},\ \bibinfo {year} {2017})\BibitemShut {NoStop}%
\bibitem [{\citenamefont {Komiske}\ \emph {et~al.}(2019)\citenamefont {Komiske}, \citenamefont {Metodiev},\ and\ \citenamefont {Thaler}}]{Komiske:2018cqr}%
  \BibitemOpen
  \bibfield  {author} {\bibinfo {author} {\bibfnamefont {P.~T.}\ \bibnamefont {Komiske}}, \bibinfo {author} {\bibfnamefont {E.~M.}\ \bibnamefont {Metodiev}},\ and\ \bibinfo {author} {\bibfnamefont {J.}~\bibnamefont {Thaler}},\ }\href {https://doi.org/10.1007/JHEP01(2019)121} {\bibfield  {journal} {\bibinfo  {journal} {JHEP}\ }\textbf {\bibinfo {volume} {01}},\ \bibinfo {pages} {121 (2019)}},\ \Eprint {https://arxiv.org/abs/1810.05165} {arXiv:1810.05165 [hep-ph]} \BibitemShut {NoStop}%
\bibitem [{\citenamefont {Qu}\ \emph {et~al.}(2022)\citenamefont {Qu}, \citenamefont {Li},\ and\ \citenamefont {Qian}}]{qu2022particle}%
  \BibitemOpen
  \bibfield  {author} {\bibinfo {author} {\bibfnamefont {H.}~\bibnamefont {Qu}}, \bibinfo {author} {\bibfnamefont {C.}~\bibnamefont {Li}},\ and\ \bibinfo {author} {\bibfnamefont {S.}~\bibnamefont {Qian}},\ }in\ \href@noop {} {\emph {\bibinfo {booktitle} {International Conference on Machine Learning}}}\ (\bibinfo {organization} {PMLR},\ \bibinfo {year} {2022})\ pp.\ \bibinfo {pages} {18281--18292}\BibitemShut {NoStop}%
\bibitem [{\citenamefont {Qu}\ and\ \citenamefont {Gouskos}(2020)}]{Qu:2019gqs}%
  \BibitemOpen
  \bibfield  {author} {\bibinfo {author} {\bibfnamefont {H.}~\bibnamefont {Qu}}\ and\ \bibinfo {author} {\bibfnamefont {L.}~\bibnamefont {Gouskos}},\ }\href {https://doi.org/10.1103/PhysRevD.101.056019} {\bibfield  {journal} {\bibinfo  {journal} {Phys. Rev. D}\ }\textbf {\bibinfo {volume} {101}},\ \bibinfo {pages} {056019} (\bibinfo {year} {2020})},\ \Eprint {https://arxiv.org/abs/1902.08570} {arXiv:1902.08570 [hep-ph]} \BibitemShut {NoStop}%
\bibitem [{\citenamefont {Vaswani}\ \emph {et~al.}(2017)\citenamefont {Vaswani}, \citenamefont {Shazeer}, \citenamefont {Parmar}, \citenamefont {Uszkoreit}, \citenamefont {Jones}, \citenamefont {Gomez}, \citenamefont {Kaiser},\ and\ \citenamefont {Polosukhin}}]{vaswani2017attention}%
  \BibitemOpen
  \bibfield  {author} {\bibinfo {author} {\bibfnamefont {A.}~\bibnamefont {Vaswani}}, \bibinfo {author} {\bibfnamefont {N.}~\bibnamefont {Shazeer}}, \bibinfo {author} {\bibfnamefont {N.}~\bibnamefont {Parmar}}, \bibinfo {author} {\bibfnamefont {J.}~\bibnamefont {Uszkoreit}}, \bibinfo {author} {\bibfnamefont {L.}~\bibnamefont {Jones}}, \bibinfo {author} {\bibfnamefont {A.~N.}\ \bibnamefont {Gomez}}, \bibinfo {author} {\bibfnamefont {L.~u.}\ \bibnamefont {Kaiser}},\ and\ \bibinfo {author} {\bibfnamefont {I.}~\bibnamefont {Polosukhin}},\ }in\ \href {https://proceedings.neurips.cc/paper_files/paper/2017/file/3f5ee243547dee91fbd053c1c4a845aa-Paper.pdf} {\emph {\bibinfo {booktitle} {Advances in Neural Information Processing Systems}}},\ Vol.~\bibinfo {volume} {30}\ (\bibinfo {year} {2017})\BibitemShut {NoStop}%
\bibitem [{\citenamefont {Dolen}\ \emph {et~al.}(2016)\citenamefont {Dolen}, \citenamefont {Harris}, \citenamefont {Marzani}, \citenamefont {Rappoccio},\ and\ \citenamefont {Tran}}]{Dolen:2016kst}%
  \BibitemOpen
  \bibfield  {author} {\bibinfo {author} {\bibfnamefont {J.}~\bibnamefont {Dolen}}, \bibinfo {author} {\bibfnamefont {P.}~\bibnamefont {Harris}}, \bibinfo {author} {\bibfnamefont {S.}~\bibnamefont {Marzani}}, \bibinfo {author} {\bibfnamefont {S.}~\bibnamefont {Rappoccio}},\ and\ \bibinfo {author} {\bibfnamefont {N.}~\bibnamefont {Tran}},\ }\href {https://doi.org/10.1007/JHEP05(2016)156} {\bibfield  {journal} {\bibinfo  {journal} {JHEP}\ }\textbf {\bibinfo {volume} {05}},\ \bibinfo {pages} {156 (2016)}},\ \Eprint {https://arxiv.org/abs/1603.00027} {arXiv:1603.00027 [hep-ph]} \BibitemShut {NoStop}%
\bibitem [{\citenamefont {Englert}\ \emph {et~al.}(2019)\citenamefont {Englert}, \citenamefont {Galler}, \citenamefont {Harris},\ and\ \citenamefont {Spannowsky}}]{Englert:2018cfo}%
  \BibitemOpen
  \bibfield  {author} {\bibinfo {author} {\bibfnamefont {C.}~\bibnamefont {Englert}}, \bibinfo {author} {\bibfnamefont {P.}~\bibnamefont {Galler}}, \bibinfo {author} {\bibfnamefont {P.}~\bibnamefont {Harris}},\ and\ \bibinfo {author} {\bibfnamefont {M.}~\bibnamefont {Spannowsky}},\ }\href {https://doi.org/10.1140/epjc/s10052-018-6511-8} {\bibfield  {journal} {\bibinfo  {journal} {Eur. Phys. J. C}\ }\textbf {\bibinfo {volume} {79}},\ \bibinfo {pages} {4} (\bibinfo {year} {2019})},\ \Eprint {https://arxiv.org/abs/1807.08763} {arXiv:1807.08763 [hep-ph]} \BibitemShut {NoStop}%
\bibitem [{\citenamefont {Kasieczka}\ and\ \citenamefont {Shih}(2020)}]{Kasieczka:2020yyl}%
  \BibitemOpen
  \bibfield  {author} {\bibinfo {author} {\bibfnamefont {G.}~\bibnamefont {Kasieczka}}\ and\ \bibinfo {author} {\bibfnamefont {D.}~\bibnamefont {Shih}},\ }\href {https://doi.org/10.1103/PhysRevLett.125.122001} {\bibfield  {journal} {\bibinfo  {journal} {Phys. Rev. Lett.}\ }\textbf {\bibinfo {volume} {125}},\ \bibinfo {pages} {122001} (\bibinfo {year} {2020})},\ \Eprint {https://arxiv.org/abs/2001.05310} {arXiv:2001.05310 [hep-ph]} \BibitemShut {NoStop}%
\bibitem [{\citenamefont {Kitouni}\ \emph {et~al.}(2020)\citenamefont {Kitouni}, \citenamefont {Nachman}, \citenamefont {Weisser},\ and\ \citenamefont {Williams}}]{Kitouni:2020xgb}%
  \BibitemOpen
  \bibfield  {author} {\bibinfo {author} {\bibfnamefont {O.}~\bibnamefont {Kitouni}}, \bibinfo {author} {\bibfnamefont {B.}~\bibnamefont {Nachman}}, \bibinfo {author} {\bibfnamefont {C.}~\bibnamefont {Weisser}},\ and\ \bibinfo {author} {\bibfnamefont {M.}~\bibnamefont {Williams}},\ }\href {https://doi.org/10.1007/JHEP04(2021)070} {\bibfield  {journal} {\bibinfo  {journal} {JHEP}\ }\textbf {\bibinfo {volume} {21}},\ \bibinfo {pages} {070 (2020)}},\ \Eprint {https://arxiv.org/abs/2010.09745} {arXiv:2010.09745 [hep-ph]} \BibitemShut {NoStop}%
\bibitem [{\citenamefont {Sun}\ \emph {et~al.}(2021)\citenamefont {Sun}, \citenamefont {Liu}, \citenamefont {Jing}, \citenamefont {Wang}, \citenamefont {Zhong},\ and\ \citenamefont {Song}}]{Sun:2020ehv}%
  \BibitemOpen
  \bibfield  {author} {\bibinfo {author} {\bibfnamefont {W.}~\bibnamefont {Sun}}, \bibinfo {author} {\bibfnamefont {T.}~\bibnamefont {Liu}}, \bibinfo {author} {\bibfnamefont {M.}~\bibnamefont {Jing}}, \bibinfo {author} {\bibfnamefont {L.}~\bibnamefont {Wang}}, \bibinfo {author} {\bibfnamefont {B.}~\bibnamefont {Zhong}},\ and\ \bibinfo {author} {\bibfnamefont {W.}~\bibnamefont {Song}},\ }\href {https://doi.org/10.1007/s11467-021-1085-6} {\bibfield  {journal} {\bibinfo  {journal} {Front. Phys. (Beijing)}\ }\textbf {\bibinfo {volume} {16}},\ \bibinfo {pages} {64501} (\bibinfo {year} {2021})},\ \Eprint {https://arxiv.org/abs/2011.07889} {arXiv:2011.07889 [hep-ex]} \BibitemShut {NoStop}%
\bibitem [{\citenamefont {Sinkus}\ and\ \citenamefont {Voss}(1997)}]{Sinkus:1996ch}%
  \BibitemOpen
  \bibfield  {author} {\bibinfo {author} {\bibfnamefont {R.}~\bibnamefont {Sinkus}}\ and\ \bibinfo {author} {\bibfnamefont {T.}~\bibnamefont {Voss}},\ }\href {https://doi.org/10.1016/S0168-9002(97)00524-X} {\bibfield  {journal} {\bibinfo  {journal} {Nucl. Instrum. Meth. A}\ }\textbf {\bibinfo {volume} {391}},\ \bibinfo {pages} {360} (\bibinfo {year} {1997})}\BibitemShut {NoStop}%
\bibitem [{\citenamefont {Srivastava}\ \emph {et~al.}(2014)\citenamefont {Srivastava}, \citenamefont {Hinton}, \citenamefont {Krizhevsky}, \citenamefont {Sutskever},\ and\ \citenamefont {Salakhutdinov}}]{Srivastava:2014kpo}%
  \BibitemOpen
  \bibfield  {author} {\bibinfo {author} {\bibfnamefont {N.}~\bibnamefont {Srivastava}}, \bibinfo {author} {\bibfnamefont {G.}~\bibnamefont {Hinton}}, \bibinfo {author} {\bibfnamefont {A.}~\bibnamefont {Krizhevsky}}, \bibinfo {author} {\bibfnamefont {I.}~\bibnamefont {Sutskever}},\ and\ \bibinfo {author} {\bibfnamefont {R.}~\bibnamefont {Salakhutdinov}},\ }\href {http://jmlr.org/papers/v15/srivastava14a.html} {\bibfield  {journal} {\bibinfo  {journal} {Journal of Machine Learning Research}\ }\textbf {\bibinfo {volume} {15}},\ \bibinfo {pages} {1929} (\bibinfo {year} {2014})}\BibitemShut {NoStop}%
\bibitem [{\citenamefont {Albrecht}\ \emph {et~al.}(1990)\citenamefont {Albrecht} \emph {et~al.}}]{ARGUS:1990hfq}%
  \BibitemOpen
  \bibfield  {author} {\bibinfo {author} {\bibfnamefont {H.}~\bibnamefont {Albrecht}} \emph {et~al.} (\bibinfo {collaboration} {ARGUS}),\ }\href {https://doi.org/10.1016/0370-2693(90)91293-K} {\bibfield  {journal} {\bibinfo  {journal} {Phys. Lett. B}\ }\textbf {\bibinfo {volume} {241}},\ \bibinfo {pages} {278} (\bibinfo {year} {1990})}\BibitemShut {NoStop}%
\bibitem [{\citenamefont {Workman}\ \emph {et~al.}(2022)\citenamefont {Workman} \emph {et~al.}}]{ParticleDataGroup:2022pth}%
  \BibitemOpen
  \bibfield  {author} {\bibinfo {author} {\bibfnamefont {R.~L.}\ \bibnamefont {Workman}} \emph {et~al.} (\bibinfo {collaboration} {Particle Data Group}),\ }\href {https://doi.org/10.1093/ptep/ptac097} {\bibfield  {journal} {\bibinfo  {journal} {PTEP}\ }\textbf {\bibinfo {volume} {2022}},\ \bibinfo {pages} {083C01} (\bibinfo {year} {2022})}\BibitemShut {NoStop}%
\bibitem [{\citenamefont {Ablikim}\ \emph {et~al.}(2022{\natexlab{d}})\citenamefont {Ablikim} \emph {et~al.}}]{BESIII:2022ahw}%
  \BibitemOpen
  \bibfield  {author} {\bibinfo {author} {\bibfnamefont {M.}~\bibnamefont {Ablikim}} \emph {et~al.} (\bibinfo {collaboration} {BESIII Collaboration}),\ }\href {https://doi.org/10.1103/PhysRevD.106.112007} {\bibfield  {journal} {\bibinfo  {journal} {Phys. Rev. D}\ }\textbf {\bibinfo {volume} {106}},\ \bibinfo {pages} {112007} (\bibinfo {year} {2022}{\natexlab{d}})},\ \Eprint {https://arxiv.org/abs/2210.09601} {arXiv:2210.09601 [hep-ex]} \BibitemShut {NoStop}%
\bibitem [{\citenamefont {Chernick}(2011)}]{chernick2011bootstrap}%
  \BibitemOpen
  \bibfield  {author} {\bibinfo {author} {\bibfnamefont {M.~R.}\ \bibnamefont {Chernick}},\ }\href@noop {} {\emph {\bibinfo {title} {Bootstrap methods: A guide for practitioners and researchers}}}\ (\bibinfo  {publisher} {John Wiley \& Sons},\ \bibinfo {year} {2011})\BibitemShut {NoStop}%
\bibitem [{\citenamefont {Ablikim}\ \emph {et~al.}(2021)\citenamefont {Ablikim} \emph {et~al.}}]{BESIII:2020kzc}%
  \BibitemOpen
  \bibfield  {author} {\bibinfo {author} {\bibfnamefont {M.}~\bibnamefont {Ablikim}} \emph {et~al.} (\bibinfo {collaboration} {BESIII Collaboration}),\ }\href {https://doi.org/10.1016/j.physletb.2021.136327} {\bibfield  {journal} {\bibinfo  {journal} {Phys. Lett. B}\ }\textbf {\bibinfo {volume} {817}},\ \bibinfo {pages} {136327} (\bibinfo {year} {2021})},\ \Eprint {https://arxiv.org/abs/2012.11106} {arXiv:2012.11106 [hep-ex]} \BibitemShut {NoStop}%
\bibitem [{\citenamefont {Ablikim}\ \emph {et~al.}(2015)\citenamefont {Ablikim} \emph {et~al.}}]{BESIII:2015qfd}%
  \BibitemOpen
  \bibfield  {author} {\bibinfo {author} {\bibfnamefont {M.}~\bibnamefont {Ablikim}} \emph {et~al.} (\bibinfo {collaboration} {BESIII Collaboration}),\ }\href {https://doi.org/10.1088/1674-1137/39/9/093001} {\bibfield  {journal} {\bibinfo  {journal} {Chin. Phys. C}\ }\textbf {\bibinfo {volume} {39}},\ \bibinfo {pages} {093001} (\bibinfo {year} {2015})},\ \Eprint {https://arxiv.org/abs/1503.03408} {arXiv:1503.03408 [hep-ex]} \BibitemShut {NoStop}%
\bibitem [{\citenamefont {Shanahan}\ \emph {et~al.}(2022)\citenamefont {Shanahan} \emph {et~al.}}]{Shanahan:2022ifi}%
  \BibitemOpen
  \bibfield  {author} {\bibinfo {author} {\bibfnamefont {P.}~\bibnamefont {Shanahan}} \emph {et~al.},\ }in\ \href@noop {} {\emph {\bibinfo {booktitle} {Proceedings of the 2021 US Community Study on the Future of Particle Physics}}}\ (\bibinfo {year} {2022})\ \Eprint {https://arxiv.org/abs/2209.07559} {arXiv:2209.07559 [physics.comp-ph]} \BibitemShut {NoStop}%
\bibitem [{\citenamefont {Chen}\ \emph {et~al.}(2022)\citenamefont {Chen}, \citenamefont {Dey}, \citenamefont {Ghosh}, \citenamefont {Kagan}, \citenamefont {Nord},\ and\ \citenamefont {Ramachandra}}]{Chen:2022pzc}%
  \BibitemOpen
  \bibfield  {author} {\bibinfo {author} {\bibfnamefont {T.~Y.}\ \bibnamefont {Chen}}, \bibinfo {author} {\bibfnamefont {B.}~\bibnamefont {Dey}}, \bibinfo {author} {\bibfnamefont {A.}~\bibnamefont {Ghosh}}, \bibinfo {author} {\bibfnamefont {M.}~\bibnamefont {Kagan}}, \bibinfo {author} {\bibfnamefont {B.}~\bibnamefont {Nord}},\ and\ \bibinfo {author} {\bibfnamefont {N.}~\bibnamefont {Ramachandra}},\ }in\ \href {https://doi.org/10.2172/1886020} {\emph {\bibinfo {booktitle} {{Proceedings of the 2021 US Community Study on the Future of Particle Physics}}}}\ (\bibinfo {year} {2022})\ \Eprint {https://arxiv.org/abs/2208.03284} {arXiv:2208.03284 [hep-ex]} \BibitemShut {NoStop}%
\bibitem [{\citenamefont {Wang}\ \emph {et~al.}(2019)\citenamefont {Wang}, \citenamefont {Sun}, \citenamefont {Liu}, \citenamefont {Sarma}, \citenamefont {Bronstein},\ and\ \citenamefont {Solomon}}]{Wang:2018nkf}%
  \BibitemOpen
  \bibfield  {author} {\bibinfo {author} {\bibfnamefont {Y.}~\bibnamefont {Wang}}, \bibinfo {author} {\bibfnamefont {Y.}~\bibnamefont {Sun}}, \bibinfo {author} {\bibfnamefont {Z.}~\bibnamefont {Liu}}, \bibinfo {author} {\bibfnamefont {S.~E.}\ \bibnamefont {Sarma}}, \bibinfo {author} {\bibfnamefont {M.~M.}\ \bibnamefont {Bronstein}},\ and\ \bibinfo {author} {\bibfnamefont {J.~M.}\ \bibnamefont {Solomon}},\ }\href {https://doi.org/10.1145/3326362} {\bibfield  {journal} {\bibinfo  {journal} {ACM Trans. Graph.}\ }\textbf {\bibinfo {volume} {38}},\ \bibinfo {pages} {1} (\bibinfo {year} {2019})}\BibitemShut {NoStop}%
\end{thebibliography}%
